\documentclass[12pt,aps,prb,preprint]{
revtex4-1}
\usepackage{amsmath}
\usepackage{amssymb}
\usepackage{tikz}
\usetikzlibrary{arrows}
\usetikzlibrary{calc,through,backgrounds}
\usepackage{euscript}
\usepackage{graphicx}
\usepackage{theorem}
{\theorembodyfont{\rmfamily}
\newtheorem{Lem}{Lemma}}
\newcommand{\eye}{\uvec{\imath}}
\newcommand{\jay}{\uvec{\jmath}}
\newcommand{\kay}{\uvec{k}}

\newcommand{\dotp}{\ensuremath{{\raise 1.5pt
\hbox{$ \pmb{.} $ }}}}
\newcommand{\mtud}[3]{\mbox{${#1}^{#2}_{\,
\;#3}$}}
\newcommand{\pvec}[1]{{\ensuremath{\vec{#1}\,
'}}}
\newcommand{\uvec}[1]{\ensuremath{\hat{#1}}}
\newcommand{\sfx}[1]{{\ensuremath{_{#1}}}}

\begin{document}

\title{On time-interval transformations in
special relativity}

\author{A.V.Gopala Rao\footnote{Formerly
Professor of Physics},
K.S.Mallesh and K.N.Srinivasa
Rao\footnote{Formerly Professor
of Physics}},
\affiliation{Department of Studies  in
Physics,  University of
Mysore, Manasagangotri,  Mysore 570 006,
INDIA}
\email{garakali@gmail.com}

\date{\today}
\hfill[Final revised version]

\begin{abstract} We revisit the problem of
the Lorentz transformation of time-intervals
in special relativity. We base our discussion
on the time-interval transformation formula
$ c\Delta t' = \gamma (c\Delta t -
\vec{\beta} \cdot \Delta \vec{r}) $ in which
$ \Delta t'$ and $ \Delta t $ are the
time-intervals between a given pair of
events, in two inertial frames $ S $ and $
S'$ connected by an general boost. We observe
that the Einstein time-dilation-formula, the
Doppler formula and the relativity of
simultaneity, all follow when one the frames
in the time-interval transformation formula
is chosen as the canonical frame of the
underlying event-pair. We also discuss the
interesting special case $ \Delta t' = \gamma
\Delta t $ of the time-interval
transformation formula obtained by setting  $
\vec{\beta} \cdot \Delta \vec{r}=0 $ in it
and argue why it is really \textbf{not} the
Einstein time-dilation formula. Finally, we
present some examples which involve material
particles instead of light rays, and
highlight the utility of time-interval
transformation formula as a calculational
tool in the class room.
\end{abstract}

\maketitle

\section{Notation and convention}
$ \mathbb{M} $ denotes the Minkowski
spacetime. We work in
signature $ +--- $. Events  in $ \mathbb{M} $
are denoted by Euler-Script characters such
as $ \EuScript{P}_1 $ and $ \EuScript{P}_2 $.
Latin suffixes are used for the space-range
1,2,3 and Greek suffixes for the spacetime
range 0,1,2,3. $ S : \{ct, x, y, z\} $ and $
S' : \{ct', x', y', z' \} $ are two inertial
coordinate systems in $ \mathbb{M} $. An
event, say $ \EuScript{P}_1 $, has the
coordinates $ (ct_1,x_1,y_1,z_1)\equiv
(ct_1,\vec{r}_1) $ in the inertial frame $ S
$ and the coordinates $
(ct'_1,x'_1,y'_1,z'_1)\equiv
(ct'_1,\pvec{r}_1) $ in the inertial frame $
S' $. The standard symbols $ \beta $ and $
\gamma $ denote $ v / c $ and $ 1/ \sqrt{1 -
v^2 / c^2} $.

\section{Lorentz transformation of
time-intervals}
The discussion in this paper is aimed at
students of physics who have had a first
course in special relativity and are familiar
with some ideas in Minkowski geometry such as
events, Lorentz invariance of spacetime
intervals, time-like, null and spacelike
intervals, and especially the general Lorentz
transformation in which the relative velocity
between the frames is not along a common
spatial axis.

We begin by recalling that an inertial frame
is essentially a set of devices which enable
setting up a pseudo-Cartesian coordinate
system covering the whole of the Minkowski
spacetime $ \mathbb{M} $. In a given inertial
frame, say $ S $, every event  $ \mathcal{P}
$ is associated with a unique quadruplet $
(ct,x,y,z) $ of pseudo-Cartesian coordinates.
Here, we recall that the time-coordinate in
an inertial frame would have been set up
using a system of synchronized standard
clocks at rest relative to $ S $.

Let $ \{\EuScript{P}_1, \EuScript{P}_2\} $ be
a given pair of events with $ S
$-frame spacetime coordinates $
(ct_1,\vec{r}_1) $ and $ (ct_2,\vec{r}_2) $.
Further, let $ \EuScript{P}_2 $ occur later
than $ \EuScript{P}_1 $ in $ S $. Then, the
displacement 4-vector between them is given
by $ (ct_2-ct_1, x_2-x_1,y_2-y_1,z_2-z_1)
\equiv (c\Delta t_\sfx{12},\Delta x_\sfx{12},
\Delta y_\sfx{12}, \Delta z_\sfx{12}) $,
where we may remember that we have
conveniently chosen $ \Delta t_\sfx{12} > 0
$ in the $ S $-frame. Events $ \EuScript{P}_1
$ and $ \EuScript{P}_2 $  are said to be
\textbf{timelike-separated},
\textbf{null-separated}, or
\textbf{spacelike-separated} according as
\begin{align*}
c^2\,\Delta t_\sfx{12}^2-\Delta
x_\sfx{12}^2-\Delta y_\sfx{12}^2-\Delta
z_\sfx{12}^2 \gtreqqless 0.
\end{align*}
Further, we have the following easily proved,
well known, results concerning  pairs of
events of $ \mathbb{M} $:
\Lem  A pair of  timelike-separated events is
\textbf{contiguous} (i.e., they occur
at the same spatial point) in an appropriate
inertial frame called the canonical inertial
frame (or, the proper-frame)  of  the
timelike-separated event-pair.
\Lem  A pair of spacelike-separated events is
\textbf{simultaneous} (i.e., they occur at
the same instant) in an appropriate canonical
inertial frame of  the spacelike-separated
event-pair.
\Lem A pair of null-separated events has
space and time separations which are related
by $ c\,\Delta t_\sfx{12} = | \Delta
\vec{r}_\sfx{12} | $ in \textbf{every}
inertial frame $ S $.

In what follows, we use the rule for
transforming the time-interval between an
(arbitrary) event-pair in one inertial frame
$ S:\{x^\mu\} $ to that in another inertial
frame  say, $ S':\{x'^\mu\} $. Since we do
not want to
restrict to any particular direction for the
relative motion  between the frames $ S $ and
$ S' $, we consider $ S $ and $ S'$ to be
connected by a \textbf{general Lorentz boost}
[see for example, Weinberg, Ref.~3,
or, Misner, Wheeler and Thorne, Ref~4.]
\begin{align}\label{1}
x'^\mu = \mtud{L} {\mu} {\nu}x^\nu,
\end{align}
where the Lorentz-matrix $ [\mtud{L}{\mu}
{\nu}] $ has the elements
\begin{align} \label{2}
\mtud{L}{0}{0}   = \gamma, \; \;
\mtud{L}{0}{i}  =
\mtud{L}{i} {0}= - \gamma \beta_{i}, \;\;
\mtud{L}{i} {j}=
\delta_{ij} + (\gamma - 1) \beta_{i}
\beta_{j} /\beta^2,
\end{align}
in which $ c \vec{\beta} = c (\beta_1\eye  +
\beta_2 \jay + \beta_3\kay
) $ is the constant 3-velocity of the
Cartesian axes\break $ X'Y'Z' $ of $ S'$
relative to the Cartesian axes $ XYZ $  of $
S $,
$ \vec{\beta} =  \beta\, \uvec{\beta} $ and $
\gamma = (1 - \beta^2)^{- 1 / 2}$.

If we write down  the zeroth components of
Eq.\eqref{1} for two arbitrary events $
\EuScript{P}_1   $ and $ \EuScript{P}_1 $
with coordinates $  (ct_1,\vec{r}_1) $ and $
(ct_2,\vec{r}_2) $ in $ S $, subtract the
relation so obtained for $ \EuScript{P}_1 $
from that of $ \EuScript{P}_1 $, we get
\begin{align} \label{3}
\Delta t'_\sfx{12}  =\gamma \left[\Delta
t_\sfx{12}-(\vec{\beta}/c)\, \,\dotp\,\Delta
\vec{r}_\sfx{12}\right].
\end{align}
This equation which may be called as the
\textbf{general time-interval transformation
formula} is our \textbf{key formula}.

In the following paragraphs, taking one of
the inertial frames in Eq.~\eqref{3}, say $ S
$, to be the canonical frame in the cases of
timelike- and spacelike- separated
event-pairs, we determine the resulting
``canonical forms''  of Eq.~\eqref{3}. In the
remaining case of the null-separated
event-pair, because the speed of light is
isotropic and has the same value in all
inertial frames, it appears that
Eq.~\eqref{3} has no special canonical form.
But we shall see that the
Doppler-frequency formula is indeed a
canonical form associated with Eq.~\eqref{3}.
These observations show how the intrinsic
nature of Eq.\eqref{3} depends on the
invariant-type of the event-pair considered.
\subsection{Canonical form of Eqn.~\eqref{3}
for a timelike-separated event-pair}
Let the frame $ S $ be a canonical frame of
the timelike-separated event-pair $
\{\mathcal{P}_1,\mathcal{P}_2 \}$. Then, in $
S $, $ \Delta \vec{r}_\sfx{12} =
\vec{r}_2-\vec{r}_1=0 $ and the two events $
\EuScript{P}_1 $  and $ \EuScript{P}_2 $ have
 a \textbf{proper time-interval} $ \Delta
t_\sfx{12} \equiv \Delta \tau_\sfx{12} >0 $
between them. Thus,  Eq.~\eqref{3} takes the
canonical form
\begin{align}\label{4}
\Delta t'_\sfx{12} =\gamma  \Delta
\tau_\sfx{12},
\end{align}
well known as the \textbf{Einstein
time-dilation formula} [Refs.2-7].
\subsection{Canonical form of Eqn.~\eqref{3}
for a null-separated event-pair}
Recall Lemma~3 of section~II for a
null-separated event-pair. It says that there
is no canonical Lorentz frame for a
null-separated event-pair. (Or, we may say
that every inertial frame is ``equally
canonical'' for a null-separated event-pair.)
So, let  $ | \Delta \vec{r}_\sfx{12} | = c\,
\Delta t_\sfx{12}>0 $, in $
S $ for a given null-separated-pair of events
$\{\EuScript{P}_1,\EuScript{P}_2\} $.  Then,
in this case, the time-interval
transformation \eqref{3} may also be written
as
\begin{align}\label{5}
\Delta t'_\sfx{12} & ={\gamma}\, \Delta
t_\sfx{12}
\left(1-{\beta}\cos{\theta}\right),
\end{align}
where $ \theta $ is the angle between the
3-vectors $ \Delta \vec{r}_\sfx{12} $ and $
\vec{\beta} $ in $ S $.

It is interesting to note the form assumed by
Eq.~\eqref{5} when it is applied to a
``light-particle'' (monochromatic light
wave) and its (material-point) source. For
the light-particle, we take
$\{\EuScript{P}_1,\EuScript{P}_2\} $ to be a
pair of null-separated events on its (null)
worldline. Although there is no canonical
frame for a light-particle (in the sense of
Lemma~3, section~II), we do have a canonical
frame (rest frame) for its (material-point)
source which we may take as the frame $ S $
in Eq.~\eqref{5}. We take the time-interval $
\Delta t_\sfx{12} \equiv T>0 $ as the
\textbf{period} of the photon (wave) (more
specifically the \textbf{proper-period}) in $
S $ in which its source is at rest. Then, $
\nu =1/T $ is the \textbf{proper-frequency}
and $ \nu' =1/T'=1/\Delta t'_\sfx{12}  $
given by Eq.~\eqref{5} is its
\textbf{relative-frequency} in the frame $ S'
$, in which the source has a uniform velocity
$ c\vec{\beta} $. Now, Eq.~\eqref{5},
re-written in terms of frequencies of the
photon in the two frames, is
\begin{align}\label{6}
\frac{\nu'}{\nu}&
=\frac{\sqrt{1-v^2/c^2}}{\left(1-{\beta}\cos{
\theta}\right)},
\end{align}
which is the \textbf{relativistic Doppler
formula} [see for example Landau and
Lifshitz, Ref.~5, pp.116-17]. Here, in
Eq.\eqref{6},  $ \theta $ is the angle
between the direction of propagation (or, the
 wave 3-vector) of the plane electromagnetic
wave and the direction of motion ($\,
\vec{\beta}\, $) of its source.

\subsection{Canonical form of Eqn.~\eqref{3}
for spacelike-separated event-pairs}
If one of the frames, say $ S $, is the
canonical frame of the two
spacelike-separated events $\{\EuScript{P}_1,
\EuScript{P}_2\} $, (see  Lemma~3 of
section~II), then we have $
\Delta t_\sfx{12} = 0 $ in $ S $ and the
time-interval transformation \eqref{3} takes
the form
\begin{align}\label{7}
\Delta t'_\sfx{12}=-\gamma (\vec{\beta} \,
\,\dotp\,\Delta \vec{r}_\sfx{12})/c=-(\gamma
\beta \,\Delta L_\sfx{12}/c) \cos{\theta},
\end{align}
where $ \Delta L_\sfx{12}=|\Delta
\vec{r}_\sfx{12}| $ is the \textbf{proper
distance (length)} between the
spacelike-separated events $ \EuScript{P}_1 $
and $ \EuScript{P}_2 $ and $ \theta $ is the
angle between $ \Delta \vec{r}_\sfx{12} $ and
$ \vec{\beta} $ in $ S $. As the
proper-distance $ \Delta L_\sfx{12} $
between an spacelike-separated event-pair is
never zero (as otherwise the two events $
\EuScript{P}_1 $ and $ \EuScript{P}_2 $ would
coincide with each other!), the above formula
\eqref{7} shows that  $ \Delta t'_\sfx{12}
\neq 0 $ in $ S' $ although  $ \Delta
t'_\sfx{12} = 0 $ in $ S $. In fact, $ \Delta
t'_\sfx{12} \gtreqqless 0 $ in $ S' $
depending on $ \cos{\theta} $. We recognize
Eq.\eqref{7} as a statement of the
\textbf{relativity of simultaneity of two
spacelike-separated events} $ \EuScript{P}_1
$ and $ \EuScript{P}_2 $.
\subsection{The transverse special case of
Eqn.~{3}}
Lastly, we consider the interesting special
case of the time-interval transformation
\eqref{3} for an arbitrary pair of events $
\EuScript{P}_1 $ and $ \EuScript{P}_2 $ for
which  $ \Delta \vec{r}_\sfx{12} \neq 0 $ in
$ S $, but the term $
\vec{\beta}\,\dotp\Delta \vec{r}_\sfx{12} = 0
$ because $ \vec{\beta} $ is perpendicular to
$ \Delta \vec{r}_\sfx{12} $. This
corresponds to the situation in which the
frame  $ S'$ moves in a direction
\textbf{transverse} (or, perpendicular) to
the space 3-vector $ \Delta \vec{r}_\sfx{12}
$ in $ S $. Then, \textbf{for all event-pairs
with $ \Delta \vec{r}_\sfx{12} \neq 0 $ in $
S $}, Eq.~{3} reduces to
\begin{align}\label{8}
\Delta t'_\sfx{12}  =\gamma\Delta t_\sfx{12}.
\end{align}
This equation \eqref{8} looks exactly like
the Einstein time-dilation formula \eqref{4}
and moreover, the two equations \eqref{8} and
\eqref{4} have the same \textbf{algebraic}
content. However, there is an important
\textbf{difference} between the two: Recall
that \textsl{of all the time-intervals
between a given timelike-separated-pair of
events (realized in all possible inertial
frames), the proper time-interval is the
shortest}. Now, consider Eq.\eqref{8} when
the two events $ \EuScript{P}_1 $ and $
\EuScript{P}_2 $ occurring in it are
timelike-separated:
\begin{itemize}

\item Then, neither the $
\Delta t_\sfx{12} (>0) $ in $ S $, because of
the assumed condition $ \Delta
\vec{r}_\sfx{12} \neq 0 $  in $ S $, nor the
$ \Delta t'_\sfx{12} =\gamma\,\Delta
t_\sfx{12} $ in $ S' $ which is greater than
$ \Delta t_\sfx{12} $, and hence is not the
minimal time-interval between the events, can
be proper time-intervals.
\item On the other
hand, when the event-pair $ \{\EuScript{P}_1,
\EuScript{P}_2\} $ is null-separated or
spacelike-separated, by definition, no
inertial frame exists in which the two events
occur at the same spatial point and hence the
time-interval between them is non-proper.
\end{itemize}
Thus, irrespective of the invariant-type of
the event-pair considered, both $ \Delta
t'_\sfx{12} $  and $ \Delta t_\sfx{12} $ in
Eq.~\eqref{8} are \textsl{non-proper
time-intervals}. In contrast, in the the
Einstein time-dilation formula~\eqref{4},  $
\Delta t_\sfx{12} \equiv \Delta\tau_\sfx{12}
$ is a proper-time-interval  whereas $ \Delta
t'_\sfx{12} $ is non-proper. For this reason,
the Einstein time-dilation formula \eqref{4}
is \textbf{not} one of the transverse
time-transformation formulas in Eq.\eqref{8}.
We may summarize this observation as follows:

\begin{quote} \textsl{The Einstein
time-dilation formula is a relation
connecting the time-interval $ \Delta
t'_\sfx{12} $ between a
timelike-separated-pair of events $
\EuScript{P}_1 , \EuScript{P}_2 $ in an
\textbf{arbitrary} inertial frame $ S' $ with
the (unique) proper-time separation $ \Delta
t_\sfx{12}\equiv \Delta\tau_\sfx{12} $
between the same pair of events realized in
the rest-frame (or, canonical-frame) $ S $ of
the events.}
\end{quote}
\subsection{Deriving the Einstein
time-dilation formula}
Many of the popular gedanken experiments that
\textsl{intend} deriving  the Einstein
time-dilation formula \eqref{4} involve
comparing the time-of-flight of a light ray
in two inertial frames. Such experiments fall
into the following two categories:
\begin{figure}
\begin{center}
\begin{tikzpicture}[>=stealth',scale=.8]
\coordinate (Pone) at (2.5,0);
\coordinate (Ptwo) at (0,2.5);
\coordinate (Ponetwo) at
($(Pone)!.5!(Ptwo)$);
\coordinate (Ptri) at (2.5,5);
\coordinate (Ptwotri) at
($(Ptwo)!.5!(Ptri)$);
\draw[very thick] (Pone)--(Ptwo);
\draw[very thick,->] (Pone)--(Ponetwo);
\draw[very thick] (Ptwo)--(Ptri);
\draw[very thick,->] (Ptwo)--(Ptwotri);
\draw[fill] (Pone) circle
(1mm) ;
\draw[fill] (Ptwo) circle (1mm) ;
\draw(2.5,0) node[right]{$\EuScript{P}_1 $} ;
\draw(0.1,2.5) node[left]{$ \EuScript{P}_2
$} ;
\draw[very thick] (0,-1)--(0,5.5);
\draw[below] (0,-1.2)node{$\vec{r}_2$} ;
\draw (1.2,0)node[above]{$ct_1$};
\draw (1.2,2.5)node[above]{$ct_2$};
\draw (1.2,5)node[above]{$ct_3$};

\draw[very thick](2.5,-1)--(2.5,5.5);
\draw[below] (2.5,-1.2) node{$\vec{r}_1$} ;
\draw[fill] (2.5,5) circle (1mm);
\draw[very thick](0,2.5)--(2.5,5);
\draw[very thick,dotted] (0,0)--(2.5,0);
\draw (2.5,5)node[right]{$\EuScript{P}_3 $} ;
\draw[very thick,dotted] (0,5)--(2.5,5);
\draw[very thick,dotted] (Ptwo)--(2.5,2.5);
\end{tikzpicture}
\end{center}
\caption{}
\end{figure}
The first of these involve a
\textbf{null-separated event-pair} $
\{\mathcal{P}_1,\,\mathcal{P}_2\} $
at the ends of a segment of the worldline
of a light ray (figure~1). As described in
the frame $ S $, event $ \mathcal{P}_1 $ is
the emission of a light ray (by a
material-point-source) at $ (t_1,\,
\vec{r}_1) $, and the event $ \mathcal{P}_2 $
is the arrival of the same light ray at  $
(t_2,\,\vec{r}_2) $ where $ \vec{r}_2\neq
\vec{r}_1 $ and $ t_2>t_1 $. The
times-of-flight of the light ray in the two
frames are then related by Eq.~\eqref{5}, or,
its special case Eq.~\eqref{8}.  As observed
earlier, experiments involving inter-frame
geometries [such as the one in Ref.~1, or,
the one on  page 486 of Ref.~2, for example,]
in which $ \Delta \vec{r}_\sfx{12} \neq 0 $,
but $ \vec{\beta} $ is perpendicular to $
\Delta \vec{r}_\sfx{12} $ in $ S $,  derive
Eq.~\eqref{8} and \textbf{not} the  Einstein
time-dilation formula Eq.~\eqref{4}.

The other category of experiments that
\textsl{succeed} in deriving  the Einstein
time-dilation\break formula~\eqref{4},
too, compare
the times-of-flight of a light ray in two
inertial frames, but, involve (figure~1) a
\textbf{timelike-separated event-pair} $
\{\mathcal{P}_1,\,\mathcal{P}_3\} $ that
occur at a single space point $ \vec{r}_1 $
in the $ S $-frame. As described in the frame
$ S $, the event $ \mathcal{P}_1 $
corresponds to the emission of a light ray
(by a material-point-source) at a spatial
point $ \vec{r}_1 $ and the event $
\mathcal{P}_2 $ to  the \textbf{return} of
that light ray, after a lapse of time, to the
same spatial point $ \vec{r}_1 $ at which it
was emitted (after being reflected at some
other spatial point $ \vec{r}_2 $). Now,
interestingly, the two events $
\EuScript{P}_1 ,\EuScript{P}_3 $ lie on the
null-worldline of the light ray $
\EuScript{P}_1\, \EuScript{P}_2\,
\EuScript{P}_3 $, as well as on  the
timelike-worldline $ \EuScript{P}_1\,
\EuScript{P}_3 $. Thus, as far as $
\EuScript{P}_1 $ and $ \EuScript{P}_3 $ are
concerned, which have $
\Delta \vec{r}_\sfx{13} =0 $ in $ S $, the
relevant formula relating the times-of-flight
of the light ray in the two frames is
Eq.~\eqref{4} which is the Einstein
time-dilation formula.

\subsection{Deriving length-contraction using
Eq.\eqref{3}}
Next, we consider \textbf{a rigid rod at
rest on the $ X' $-axis} of the inertial
frame $ S' $ with a point-lamp fixed at one
end. A light ray leaves the point-lamp at the
end $ \pvec{r}_1 = (x'_1,0,0) $ of the rod at
the instant $ t'_1 $ and reaches its other
end $ \pvec{r}_2 =(x'_2>x'_1,0,0) $ at the
instant $ t'_2 $. The associated events are $
\EuScript{P}_1 =(ct_1',\pvec{r}_1) $ and $
\EuScript{P}_2 =(ct_2',\pvec{r}_2) $.
Evidently $ L_0 \equiv L'=|\pvec{r}_2-
\pvec{r}_1| =x'_2-x'_1 $ is the
\textbf{proper length of the rod}. Since the
rod is at rest and has a length $ L'=L_0 $ in
the frame $ S' $, the events  $
\EuScript{P}_1 $ and $ \EuScript{P}_2 $ are
evidently separated in time by $ \Delta
t'_\sfx{12} = L_0/c $ in $ S' $.

Note that the rigid rod moves with the
velocity $ \eye\beta/c =\eye v $ relative to
$ S $. Therefore, in $ S $, in the
time-interval $ \Delta t_\sfx{12} =(t_2-t_1)
$, the mirror-end of the rod moves through
the distance $ v\, \Delta t_\sfx{12} $ while
the light ray travels the distance $ c\,
\Delta t_\sfx{12} $. Thus, $ c\, \Delta
t_\sfx{12}=L+v\,\Delta t_\sfx{1 2} $ where $
L $ is the \textbf{length of the (moving)
rod} in the frame $ S $ and we get $ \Delta
t_\sfx{12}=L/(c-v) $.

Using Eq.~\eqref{5} for the events $
\EuScript{P}_1 $ and $ \EuScript{P}_2 $, we
get, as $ \theta=0 $ now,
\begin{align}\label{12}
\Delta t'_{12} = \gamma(1 -\beta)\, \Delta
t_{12}.
\end{align}
Then, if we plug in  $ \Delta t_{12} = \Delta
L / (c - v) $ and $ \Delta t'_{1 2} = \Delta
L_0 / c \;$ in this equation, we get $ \Delta
L_0 / c = \gamma (1 - \beta)\, \Delta L / (c
- v) $ so that $ \Delta L_0 = \gamma\, \Delta
L $ which is precisely the Lorentz-Fitzgerald
length-contraction formula.
\subsection{A different gedanken experiment
to derive length-contraction} This experiment
is essentially the same as the one described
in the previous paragraph~F with one change:
It uses a material particle (such as a bullet
shot from a gun) in the place of the light
ray. We have included this example to demonstrate
that Eq.~\eqref{3} is a good starting point to such
calculations. Moreover, this example also shows that
material particles can as well be used in the
place of light rays in such gedanken
experiments--a point which we believe, is worth bringing
to the notice of a class-room in relativity.
The \textsl{disadvantage}, however, is
that the calculations now become a little
clumsy in view of the fact that the speed of
a material particle, unlike $ c $, changes
from frame to frame.

In this calculation, it is convenient to
consider the inverse of Eq.~\eqref{3}, namely
\begin{align}\label{12a}
\Delta t_\sfx{12} =\gamma[\Delta
t'_\sfx{12}+(\vec{\beta}/c)\dotp\Delta
\pvec{r}_\sfx{12}]
\end{align}
which is obtained by changing  $ \vec{\beta}
$ to $ -\vec{\beta} $ in Eq.~\eqref{3} and
rearranging. Further, we consider the
inter-frame configuration $ \beta_1 = \beta =
v/c,\beta_2 = \beta_3=0 $ here. The gendanken
experiment is as follows: In its
\textbf{rest-frame} $ S' $, let the two ends
of a rigid-rod be at $ \vec{r}_1=(x'_1,0,0) $
 and  $ \vec{r}_2=(x'_2,0,0) $. Then, $
L'=|\vec{r}_2-\vec{r}_1|\equiv L_0=x'_2-x'_1
$ is the \textbf{proper-length of the rod}.
Let a bullet shot from a gun at $ \vec{r}_1 $
at the instant $ t'_1 $, travel with the
uniform velocity $ \eye'\,u' $ and reach  $
\vec{r}_2 $ at time $ t'_2 $. This trip of
the bullet defines the two events $
\EuScript{P}_1 $ and  $ \EuScript{P}_1  $,
with $ S' $-frame coordinates $
(ct'_1,\vec{r}_1) $ and $ (ct'_2,\vec{r}_2)
$. Let the same two events have the
$ S $-frame coordinates $ (ct_1,x_1,0,0)$
and $ (ct_2,x_2,0,0) $. Then, from
Eq.\eqref{12a}, we get $ \Delta
t_\sfx{12}=\gamma[\Delta
t'_\sfx{12}+(\beta/c)\,\Delta
x'_\sfx{12}] $, where $ \Delta
x'_\sfx{12}=x'_2-x'_1 = L'=L_0 $ and $
\Delta t'_\sfx{12} =L'/u' =L_0/u' $. We also
note that  $ \Delta t_\sfx{12}= L/(u-v) $.
Thus,
\begin{align}\label{14}
\Delta t_\sfx{12}=L/(u-v)=L_0\gamma[
1/u'+(v/c^2)].
\end{align}
Now, using the Einstein velocity addition
formula  $ u'=(u-v)/(1-vu/c^2)
$, we rewrite the above equation as $
L/L_0\gamma=[(1-vu/c^2)/(u-v)
+(v/c^2)](u-v) $, which simplifies to $
L/L_0\gamma = 1-vu/c^2 +
(u-v)v/c^2=1-v^2/c^2 = 1/\gamma^2 $, so that
$ L=L_0\gamma $ which is, again,the
length-contraction formula.

In passing, we note that two variants of the
above gedanken experiment can be tried out
for fun. In the first, we may use a material
particle doing a round trip along the $ x
$-axis of the inertial frame $ S' $ instead
of it doing a one-way trip as done above.
Alternatively, one may consider a material
particle doing a one-way trip in the
transverse configuration (for example, along
the $ y $-axis of the inertial frame $ S' $).
We leave the details to the interested
reader.

\end{document}